\documentclass[dvips]{article}
\usepackage{supertabular,lscape,epsfig}
\usepackage{amssymb}
\usepackage{amsmath}
 
\DeclareSymbolFont{ppa}{OT1}{ppl}{m}{it}
\DeclareMathSymbol{\vv}{\mathalpha}{ppa}{'166}
 
\begin{document}
 
\newcommand{\dd}{\,{\rm d}}
\newcommand{\ie}{{\it i.e.},\,}
\newcommand{\etal}{{\it et al.\ }}
\newcommand{\eg}{{\it e.g.},\,}
\newcommand{\cf}{{\it cf.\ }}
\newcommand{\vs}{{\it vs.\ }}
\newcommand{\zdot}{\makebox[0pt][l]{.}}
\newcommand{\up}[1]{\ifmmode^{\rm #1}\else$^{\rm #1}$\fi}
\newcommand{\dn}[1]{\ifmmode_{\rm #1}\else$_{\rm #1}$\fi}
\newcommand{\upd}{\up{d}}
\newcommand{\uph}{\up{h}}
\newcommand{\upm}{\up{m}}
\newcommand{\ups}{\up{s}}
\newcommand{\arcd}{\ifmmode^{\circ}\else$^{\circ}$\fi}
\newcommand{\arcm}{\ifmmode{'}\else$'$\fi}
\newcommand{\arcs}{\ifmmode{''}\else$''$\fi}
\newcommand{\MS}{{\rm M}\ifmmode_{\odot}\else$_{\odot}$\fi}
\newcommand{\RS}{{\rm R}\ifmmode_{\odot}\else$_{\odot}$\fi}
\newcommand{\LS}{{\rm L}\ifmmode_{\odot}\else$_{\odot}$\fi}
 
\newcommand{\Abstract}[2]{{\footnotesize\begin{center}ABSTRACT\end{center}

\vspace{1mm}\par#1\par
\noindent
{~}{\it #2}}}
 
\newcommand{\TabCap}[2]{\begin{center}\parbox[t]{#1}{\begin{center}
  \small {\spaceskip 2pt plus 1pt minus 1pt T a b l e}
  \refstepcounter{table}\thetable \\[2mm]
  \footnotesize #2 \end{center}}\end{center}}
 
\newcommand{\TableSep}[2]{\begin{table}[p]\vspace{#1}
\TabCap{#2}\end{table}}
 
\newcommand{\FigCap}[1]{\footnotesize\par\noindent Fig.\  %
  \refstepcounter{figure}\thefigure. #1\par}
 
\newcommand{\TableFont}{\footnotesize}
\newcommand{\TableFontIt}{\ttit}
\newcommand{\SetTableFont}[1]{\renewcommand{\TableFont}{#1}}
 
\newcommand{\MakeTable}[4]{\begin{table}[htb]\TabCap{#2}{#3}
  \begin{center} \TableFont \begin{tabular}{#1} #4
  \end{tabular}\end{center}\end{table}}
 
\newcommand{\MakeTableSep}[4]{\begin{table}[p]\TabCap{#2}{#3}
  \begin{center} \TableFont \begin{tabular}{#1} #4
  \end{tabular}\end{center}\end{table}}

\newenvironment{references}%
{
\footnotesize \frenchspacing
\renewcommand{\thesection}{}
\renewcommand{\in}{{\rm in }}
\renewcommand{\AA}{Astron.\ Astrophys.}
\newcommand{\AAS}{Astron.~Astrophys.~Suppl.~Ser.}
\newcommand{\ApJ}{Astrophys.\ J.}
\newcommand{\ApJS}{Astrophys.\ J.~Suppl.~Ser.}
\newcommand{\ApJL}{Astrophys.\ J.~Letters}
\newcommand{\AJ}{Astron.\ J.}
\newcommand{\IBVS}{IBVS}
\newcommand{\PASP}{P.A.S.P.}
\newcommand{\Acta}{Acta Astron.}
\newcommand{\MNRAS}{MNRAS}
\renewcommand{\and}{{\rm and }}
\section{{\rm REFERENCES}}
\sloppy \hyphenpenalty10000
\begin{list}{}{\leftmargin1cm\listparindent-1cm
\itemindent\listparindent\parsep0pt\itemsep0pt}}%
{\end{list}\vspace{2mm}}

\def\TYLDA{~}
\newlength{\DW}
\settowidth{\DW}{0}
\newcommand{\dw}{\hspace{\DW}}
 
\newcommand{\refitem}[5]{\item[]{#1} #2%
\def\REFARG{#3}\ifx\REFARG\TYLDA\else, {\it#3}\fi
\def\REFARG{#4}\ifx\REFARG\TYLDA\else, {\bf#4}\fi
\def\REFARG{#5}\ifx\REFARG\TYLDA\else, {#5}\fi.}
 
\newcommand{\Section}[1]{\section{#1}}
\newcommand{\Subsection}[1]{\subsection{#1}}
\newcommand{\Acknow}[1]{\par\vspace{5mm}{\bf Acknowledgements.} #1}
\pagestyle{myheadings}
 
\newfont{\bb}{ptmbi8t at 12pt}
\newcommand{\xrule}{\rule{0pt}{2.5ex}}
\newcommand{\xxrule}{\rule[-1.8ex]{0pt}{4.5ex}}
\def\thefootnote{\fnsymbol{footnote}}

\begin{center}

{\Large\bf The Optical Gravitational Lensing Experiment\footnote{Based on 
observations obtained with the 1.3-m Warsaw telescope at Las Campanas 
Observatory of the Carnegie Institution of Washington.}.\\ 
Difference Image Analysis of LMC and SMC Data. The Catalog} 
\vskip1cm
{\bf K.~~\.Z~e~b~r~u~\'n$^{1,2}$, ~~I.~~S~o~s~z~y~\'n~s~k~i$^{1,2}$,~~
P.R.~~W~o~\'z~n~i~a~k$^{4,2}$, ~~A.~~U~d~a~l~s~k~i$^{1}$, 
~~M.~~K~u~b~i~a~k$^{1}$, 
~~M.~~S~z~y~m~a~\'n~s~k~i$^{1}$, ~~G.~~P~i~e~t~r~z~y~\'n~s~k~i$^{3,1}$, 
~~O.~~S~z~e~w~c~z~y~k$^{1}$ ~~and~~ \L.~~W~y~r~z~y~k~o~w~s~k~i$^{1}$}
\vskip3mm
{$^1$Warsaw University Observatory, Al.~Ujazdowskie~4, 00-478~Warszawa, Poland\\
e-mail: (zebrun,soszynsk,udalski,mk,msz,pietrzyn,szewczyk,wyrzykow)@astrouw.edu.pl\\
$^2$Princeton University Observatory, Princeton, NJ 08544--1001, USA\\
$^3$ Universidad de Concepci{\'o}n, Departamento de Fisica, Casilla 160--C, Concepci{\'o}n, Chile\\
$^4$ Los Alamos National Observatory, MS-D436, Los Alamos NM 85745, USA\\
e-mail: wozniak@lanl.gov}
\end{center}

\vspace{0.5cm}

\Abstract{We present the first edition of a catalog of variable stars found in 
the Magellanic Clouds using OGLE-II data obtained during four years: 
1997--2000. The catalog covers about 7 square degrees of the sky -- 21 fields 
in the Large Magellanic Cloud and 11 fields in the Small Magellanic Cloud. All 
variables were found with the Difference Image Analysis (DIA) software. The 
catalog is divided into two sections. The DC section contains FITS reference 
images (obtained by co-adding 20 best frames for each field) and profile 
photometry ({\sc DoPhot}) of all variable stars on those images. The AC 
section contains flux variations and magnitudes of detected variable stars 
obtained with DIA as well as with {\sc DoPhot}. The errors of magnitude 
measurements are 0.005~mag for the brightest stars (${I<16}$~mag) then grow to 
0.08~mag at 19~mag stars and to 0.3~mag at 20.5~mag. Typically, there are 
about 400 {\it I}-band data points and about 30 {\it V} and {\it B}-band data 
points for more than 68~000 variables. The stars with high proper motions were 
excluded from this catalog and will be presented in a separate paper. A 
detailed analysis and classification of variable stars will be presented 
elsewhere. The catalog is available in electronic form {\it via} FTP and 
through WWW interface from the OGLE Internet archive. The FTP catalog contains 
approximately 2~GB of data.}{Magellanic Clouds -- Catalogs -- Stars: 
variables: general} 

\Section{Introduction}
The second phase of the Optical Gravitational Lensing Experiment (OGLE-II) 
spans four years, from 1997 to 2000. The main goal of the project was the 
search for microlensing events, but a natural by-product after four years of 
observations is a huge database of photometric measurements for millions 
of objects. Currently, the third phase of the project (OGLE-III) is underway. 
In this paper we present the catalog of variable stars in the Magellanic 
Clouds found in the already closed dataset of OGLE-II. 

In previous papers the OGLE collaboration published {\it BVI} Maps of the 
Small Magellanic Cloud (Udalski \etal 1998a) and the Large Magellanic Cloud 
(Udalski \etal 2000), catalogs of Cepheids in the LMC (Udalski \etal 1999a) 
and SMC (Udalski \etal 1999b) and catalog of Eclipsing Binary Stars in the SMC 
(Udalski \etal 1998b). 
 
This paper describes the catalog of all variable objects found in the 
Magellanic Clouds with the Difference Image Analysis (DIA) package -- an 
implementation of Alard and Lupton (1998) optimal Point Spread Function (PSF) 
matching algorithm (Wo\'zniak 2000). The stars presented in the catalog 
represent numerous types of variability. The full details of the DIA software 
can be found in papers by Wo\'zniak (2000, hereafter Paper~I) and by 
\.Zebru\'n, Soszy\'nski and Wo\'zniak (2001, hereafter Paper~II). We 
emphasize, that this is the first, preliminary, edition of the catalog, which 
may contain some spurious variables, and some genuine variables might have 
been missed. 

\Section{Observational Data}
The photometric data were collected with the 1.3-m Warsaw telescope located at 
the Las Campanas Observatory, Chile. The telescope was equipped with the 
"first generation" camera with the SITe ${2048\times2048}$ CCD detector 
working in the driftscan mode. The pixel size was 24~$\mu$m giving the 0.417 
arcsec/pixel scale. Observations of the LMC were performed in the "slow" 
reading mode of the CCD detector with the gain 3.8~e$^-$/ADU and readout noise 
about 5.4~e$^-$. Details of the instrumentation setup can be found in Udalski, 
Kubiak and Szyma\'nski (1997). 

Regular observations of the LMC fields started on January 6, 1997, while 
observations of the SMC started on June 26, 1997. About 4.5 square degrees 
of central parts of the LMC and about 2.4 square degrees of the SMC were 
observed during four seasons. In this catalog we present data collected up to 
the end of May 2000. The DIA photometry is based on the {\it I}-band 
observations. The total number of photometric measurements for about 
$2\times10^7$ stars exceeded $6\times10^9$. The mean seeing of the collected 
data is 1\zdot\arcs34. Fig.~1. shows the~histogram of~the~seeing data. 
\begin{figure}[htb]
\vglue-4mm
\centerline{\includegraphics[width=10.5cm]{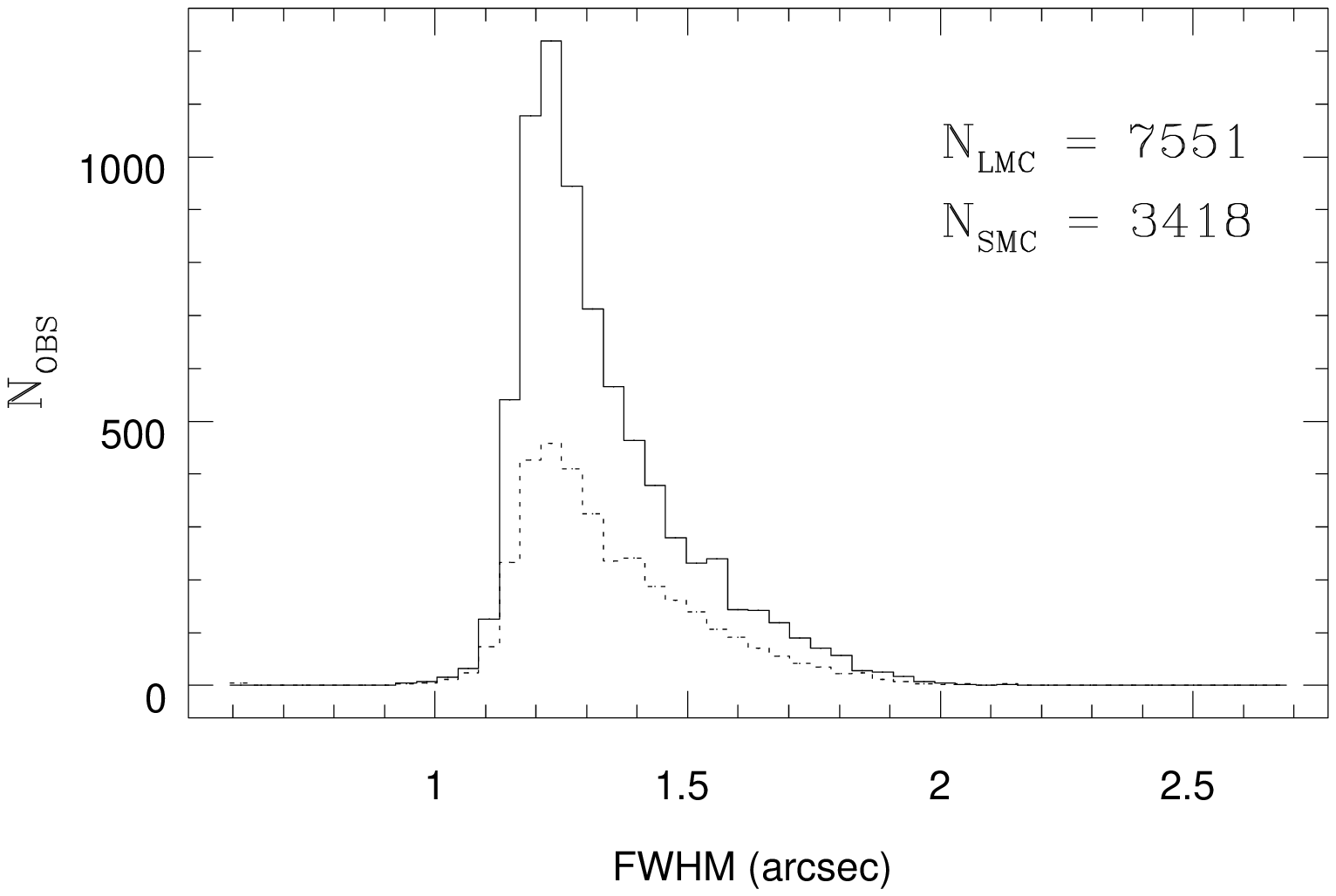}}
\FigCap{The histogram of seeing of all individual OGLE II observations of the 
Magellanic Clouds. 7551 observations of 21 LMC fields (solid line) and 3418 
observations of 11 SMC fields (dashed line) were used. The bin size is 0.041 arcsec,
corresponding to 0.1 pixel on CCD image. The mean FWHM value is 1\zdot\arcs34 
arcsec for both the LMC and SMC.} 
\end{figure}

The second edition of the catalog will include observations collected up to 
the end of November 2000, when OGLE-II phase was concluded. 

\Section{The Catalog}
There is a trade-off between the amount of time allocated to the preparation 
of a catalog and the quality of the product. Given practical limitations we 
decided to present a catalog which is preliminary rather than final. We are 
just entering the era of Tera Byte sized datasets, and it seems that results 
of such big projects will take an evolving form. We intend to correct in 
the future editions whatever deficiencies will be found in the current catalog. 

Because the dataset obtained with the DIA method is very large, we decided to 
make this data available only in the electronic form. Below we describe 
contents of our catalog, and in the next Section we present brief instructions 
on how to use the catalog. The catalog contains more than 68~000 {\it I}-band 
light curves for variable objects found among almost ${17\times10^6}$ objects 
detected in the LMC and SMC. The magnitudes of stars are transformed to the 
standard system (Udalski \etal 2000). The errors of magnitude measurements are 
0.005~mag for the brightest stars (${I<16}$~mag) then grow to 0.08~mag at
19~mag stars and to 0.3~mag at 20.5~mag. 

\renewcommand{\arraystretch}{0.9}
\MakeTable{lrrrcc}{7cm}{LMC and SMC fields observed by OGLE-II}
{
\hline
\noalign{\vskip2pt}
Field & \multicolumn{1}{c}{$~~N_{\rm obs}$} & 
\multicolumn{1}{c}{$~~N_{\rm var}$} & 
\multicolumn{1}{c}{$~~N_{\rm total}$} & RA & DEC \\
\noalign{\vskip2pt}
\hline
\noalign{\vskip2pt}
LMC$\_$SC1  & 353 & 2146 & 633666 & 5\uph33\upm49\ups& $-70\arcd06\arcm10\arcs$\\
LMC$\_$SC2  & 512 & 4620 & 702224 & 5\uph31\upm17\ups& $-69\arcd51\arcm55\arcs$\\
LMC$\_$SC3  & 505 & 3993 & 738434 & 5\uph28\upm48\ups& $-69\arcd48\arcm05\arcs$\\
LMC$\_$SC4  & 499 & 4128 & 781083 & 5\uph26\upm18\ups& $-69\arcd48\arcm05\arcs$\\
LMC$\_$SC5  & 488 & 4032 & 760089 & 5\uph23\upm48\ups& $-69\arcd41\arcm05\arcs$\\
LMC$\_$SC6  & 483 & 3964 & 785450 & 5\uph21\upm18\ups& $-69\arcd37\arcm10\arcs$\\
LMC$\_$SC7  & 475 & 4060 & 767142 & 5\uph18\upm48\ups& $-69\arcd24\arcm10\arcs$\\
LMC$\_$SC8  & 366 & 3277 & 706370 & 5\uph16\upm18\ups& $-69\arcd19\arcm15\arcs$\\
LMC$\_$SC9  & 334 & 2774 & 655470 & 5\uph13\upm48\ups& $-69\arcd14\arcm05\arcs$\\
LMC$\_$SC10 & 333 & 2250 & 596548 & 5\uph11\upm16\ups& $-69\arcd09\arcm15\arcs$\\
LMC$\_$SC11 & 272 & 1920 & 579398 & 5\uph08\upm41\ups& $-69\arcd10\arcm05\arcs$\\
LMC$\_$SC12 & 325 & 1558 & 488292 & 5\uph06\upm16\ups& $-69\arcd38\arcm20\arcs$\\
LMC$\_$SC13 & 268 & 2729 & 551847 & 5\uph06\upm14\ups& $-68\arcd43\arcm30\arcs$\\
LMC$\_$SC14 & 268 & 1494 & 478055 & 5\uph03\upm49\ups& $-69\arcd04\arcm45\arcs$\\
LMC$\_$SC15 & 275 & 1535 & 455854 & 5\uph01\upm17\ups& $-69\arcd04\arcm45\arcs$\\
LMC$\_$SC16 & 270 & 1813 & 541564 & 5\uph36\upm18\ups& $-70\arcd09\arcm40\arcs$\\
LMC$\_$SC17 & 262 & 1523 & 484820 & 5\uph38\upm48\ups& $-70\arcd16\arcm45\arcs$\\
LMC$\_$SC18 & 268 & 1371 & 430116 & 5\uph41\upm18\ups& $-70\arcd24\arcm50\arcs$\\
LMC$\_$SC19 & 260 & 1068 & 420263 & 5\uph43\upm48\ups& $-70\arcd34\arcm45\arcs$\\
LMC$\_$SC20 & 261 & 1442 & 421199 & 5\uph46\upm18\ups& $-70\arcd44\arcm50\arcs$\\
LMC$\_$SC21 & 287 & 1417 & 447731 & 5\uph21\upm14\ups& $-70\arcd33\arcm20\arcs$\\
\hline
\noalign{\vskip2pt}
SMC$\_$SC1  & 293 & 716  & 241836 & 0\uph37\upm51\ups& $-73\arcd29\arcm40\arcs$\\
SMC$\_$SC2  & 283 & 1002 & 300524 & 0\uph40\upm53\ups& $-73\arcd17\arcm30\arcs$\\
SMC$\_$SC3  & 276 & 1476 & 398790 & 0\uph43\upm58\ups& $-73\arcd12\arcm30\arcs$\\
SMC$\_$SC4  & 299 & 1770 & 426827 & 0\uph46\upm59\ups& $-73\arcd07\arcm30\arcs$\\
SMC$\_$SC5  & 313 & 2243 & 510928 & 0\uph50\upm01\ups& $-73\arcd08\arcm45\arcs$\\
SMC$\_$SC6  & 308 & 2190 & 519334 & 0\uph53\upm01\ups& $-72\arcd58\arcm40\arcs$\\
SMC$\_$SC7  & 273 & 1691 & 448674 & 0\uph56\upm00\ups& $-72\arcd53\arcm35\arcs$\\
SMC$\_$SC8  & 285 & 1317 & 383912 & 0\uph58\upm58\ups& $-72\arcd39\arcm30\arcs$\\
SMC$\_$SC9  & 279 & 1073 & 328343 & 1\uph01\upm55\ups& $-72\arcd32\arcm35\arcs$\\
SMC$\_$SC10 & 273 & 665  & 274129 & 1\uph04\upm51\ups& $-72\arcd24\arcm45\arcs$\\
SMC$\_$SC11 & 269 & 937  & 243914 & 1\uph07\upm45\ups& $-72\arcd39\arcm30\arcs$\\
\hline
\noalign{\vskip2pt}
Total:& 10515 & 68194 & 16502826 & & \\
\hline
}

Table~1 presents a summary of observations, which we used to create the 
catalog. The consecutive columns give the field name, the number of 
{\it I}-band measurements $N_I$, the number of variable stars $N_{\rm var}$, 
the total number of stars found on reference images $N_{\rm total}$ by 
standard {\sc DoPhot} routine, and the equatorial coordinates of the field 
centers for the epoch 2000. The details of~the~method used to create reference 
frames and to obtain photometry presented in this catalog can be found in 
Paper~I and Paper~II. 

The DIA package measures separately DC (constant signal) flux and AC (variable 
signal) flux for every variable object. Therefore we decided to divide the 
catalog into two parts, DC and AC, which can be accessed independently. In the 
DC part we included the FITS files of all {\it I}-band reference images for 21 
LMC and 11 SMC fields. For each field there is a single ${2048\times8192}$ 
image of whole field and 256 images of ${512\times128}$ subframes as well. 
This partitioning of the original image and the specific size of the subframes 
reflects the fact that we had to minimize PSF variations along whole image 
(see Paper~I). 

One can also find the results of {\sc DoPhot} photometry run on the DIA reference 
image (DIA {\sc DoPhot}) and the file with zero points for each subframe. 
Transformation between the DIA flux and {\sc DoPhot} magnitudes gives only 
the instrumental magnitude, which has to be shifted to match the standard 
photometric system. Note that files with DIA {\sc DoPhot} photometry are not zero 
point corrected. The format of a {\sc DoPhot} file is shown below: 
\begin{verbatim}
           2  301.48   15.24 -12.370 0.007 409.595
           3  183.83   77.00 -12.932 0.007 401.944
           4  201.42   88.27 -12.184 0.007 407.488
           5   59.35   11.97 -11.724 0.005 426.509
\end{verbatim}

The columns mean: star number, $x$ coordinate, $y$ coordinate, 
magnitude, error, background. One must remember that the DIA {\sc DoPhot} 
photometry is obtained on a image that is a sum of twenty images. This 
photometry is intended to be used only as reference for DIA. The star $x$ 
coordinate ranges from 0 to 2048 pixels while $y$ coordinate -- from 0 to 8192 
pixels. 

The zero point file is presented in the format: $X$ section number, $Y$
section number, zero point value, \ie:
\begin{verbatim}
                        1  1   26.685
                        2  1   26.669
                        3  1   26.659
\end{verbatim}

The sections number range from 1 to 4 in $x$ coordinate and from 1 to 64 in 
$y$ coordinate.  

We also provide a table with a single row for any variable object in the 
following sequence: 
\vspace*{-6pt}
\begin{itemize}
\itemsep=-4pt
\item name of the variable, we are using convention OGLE{\it 
hhmmss.ss-ddmmss.s} \eg: OGLE050129.81--683647.0, where the name gives the 
coordinates: ${\rm RA=05\uph01\upm29\zdot\ups81}$, ${\rm DEC=-
68\arcd36\arcm47\zdot\arcs0}$, 
\item RA and DEC coordinates, these coordinates are written as\\ 
{\it hh:mm:ss.ss 
dd:mm:ss.s} and as decimal values, 
\item the $x$ and $y$ pixel coordinates of a star on the reference frame; 
these are coordinates produced by DIA package, they refer to the position 
of a variable object and therefore may be somewhat different from the position 
of a star detected by {\sc DoPhot} on the reference frame, 
\item the $X$ and $Y$ number of the subsection containing the star, 
\item DIA profile and aperture photometry with errors,
\item the number of {\it I}-band frames used in DIA, \ie the total number of 
{\it I}-band OGLE observations of a given object, 
\item the number of {\it I}, {\it V} and {\it B}-band data from OGLE databases,
\item the data for the closest star identified by {\sc DoPhot} on the DIA 
reference frame -- number of the star given by {\sc DoPhot}, distance to this 
star (pixels), its magnitude and error, 
\item the data for the closest star identified on the OGLE template -- number 
of the star in the OGLE database, distance to this star (pixels), its mean 
magnitude and error, 
\item the last column contains additional flags and remarks.
\end{itemize}

Below there is a sample table row. Because of its length, it is presented as 
three lines. 
\begin{verbatim}
 OGLE052957.63-702005.8 5:29:57.63 -70:20:05.8 5.499342 70.334944 
 60.154 11.775 1 1 4530.2812 21.5921 5251.5347 21.091 302
 282 49 0 5 0.827 15.071 0.005 20 0.819 15.124 0.058 0
\end{verbatim}

The data included in the above list are also presented in the WWW part of the 
on-line catalog. 

The AC catalog contains the DIA data for variable objects. The description of 
the databases can be found in Paper~I and Paper~II. This set of variable 
objects contains stars presenting numerous types of variability, \ie 
pulsating, eclipsing, flare, RCB stars, etc. Several examples of light curves 
are shown in Figs.~2, 3 and 4. At this time we do not attempt to classify the 
types of these variables. 
\begin{figure}[htb]
\vglue-5mm
\centerline{\includegraphics[width=13cm]{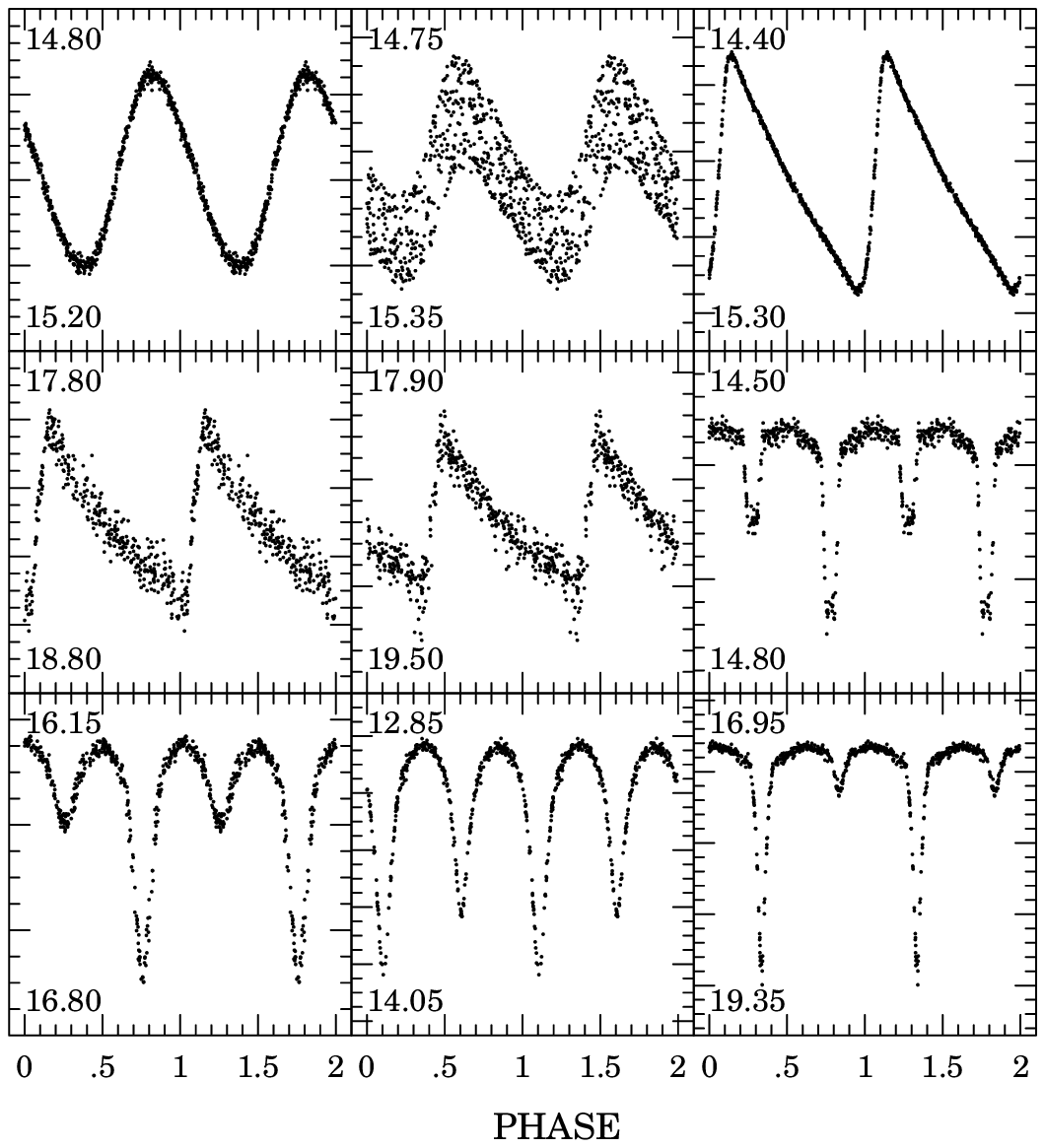}}
\FigCap{Examples of DIA light curves of the LMC$\_$SC2 pulsating and eclipsing 
stars.} 
\end{figure}
\begin{figure}[p]
\vglue-20mm
\centerline{\hglue5mm\includegraphics[width=12.5cm]{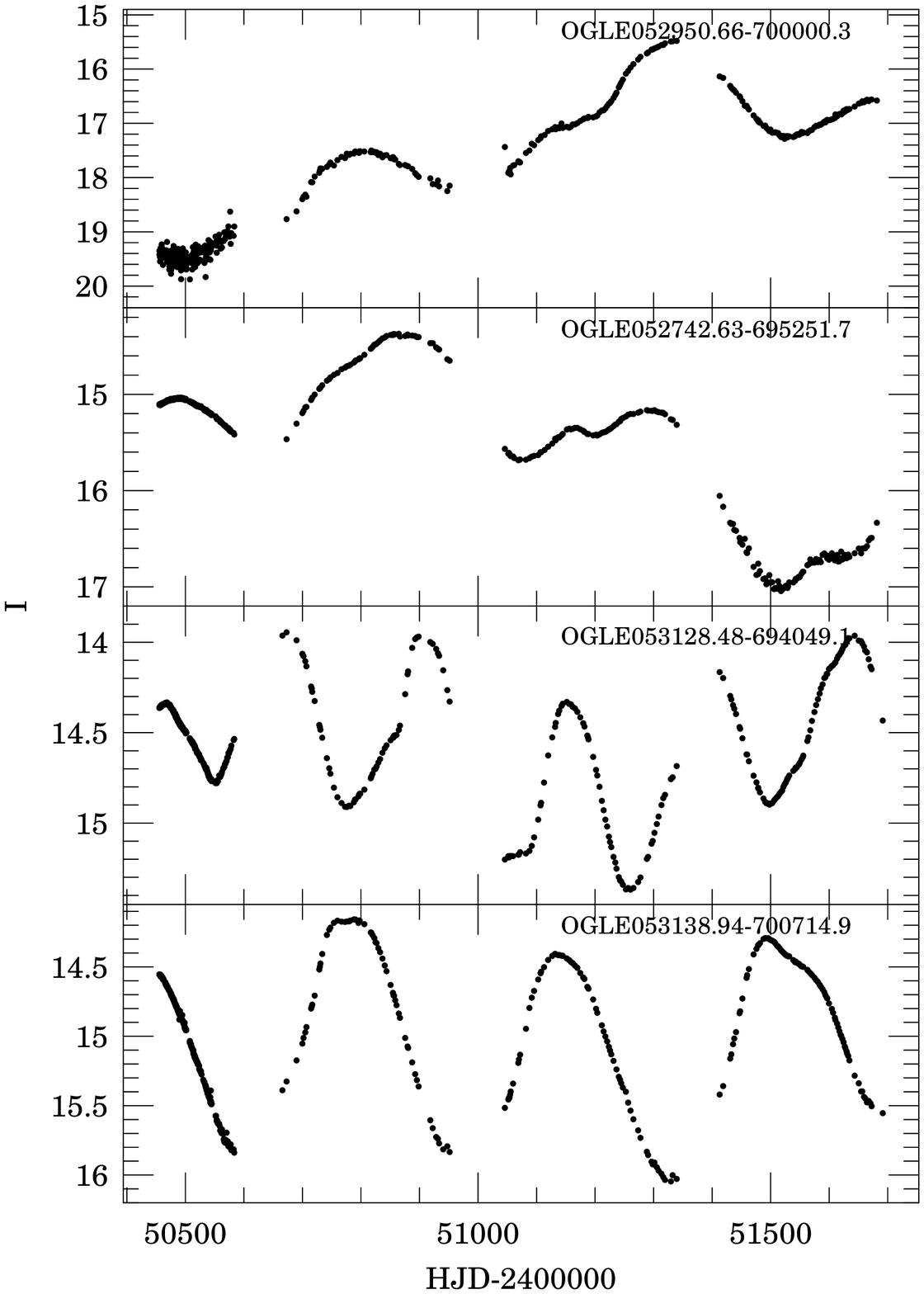}}
\vglue-10mm
\FigCap{Light curves of long period variables. The stars in the two upper 
panels are likely LPVs, while the stars in the two lower panels are good 
candidates for Mira variables.} 
\vskip4mm
\centerline{\hglue-10mm\includegraphics[width=9.7cm]{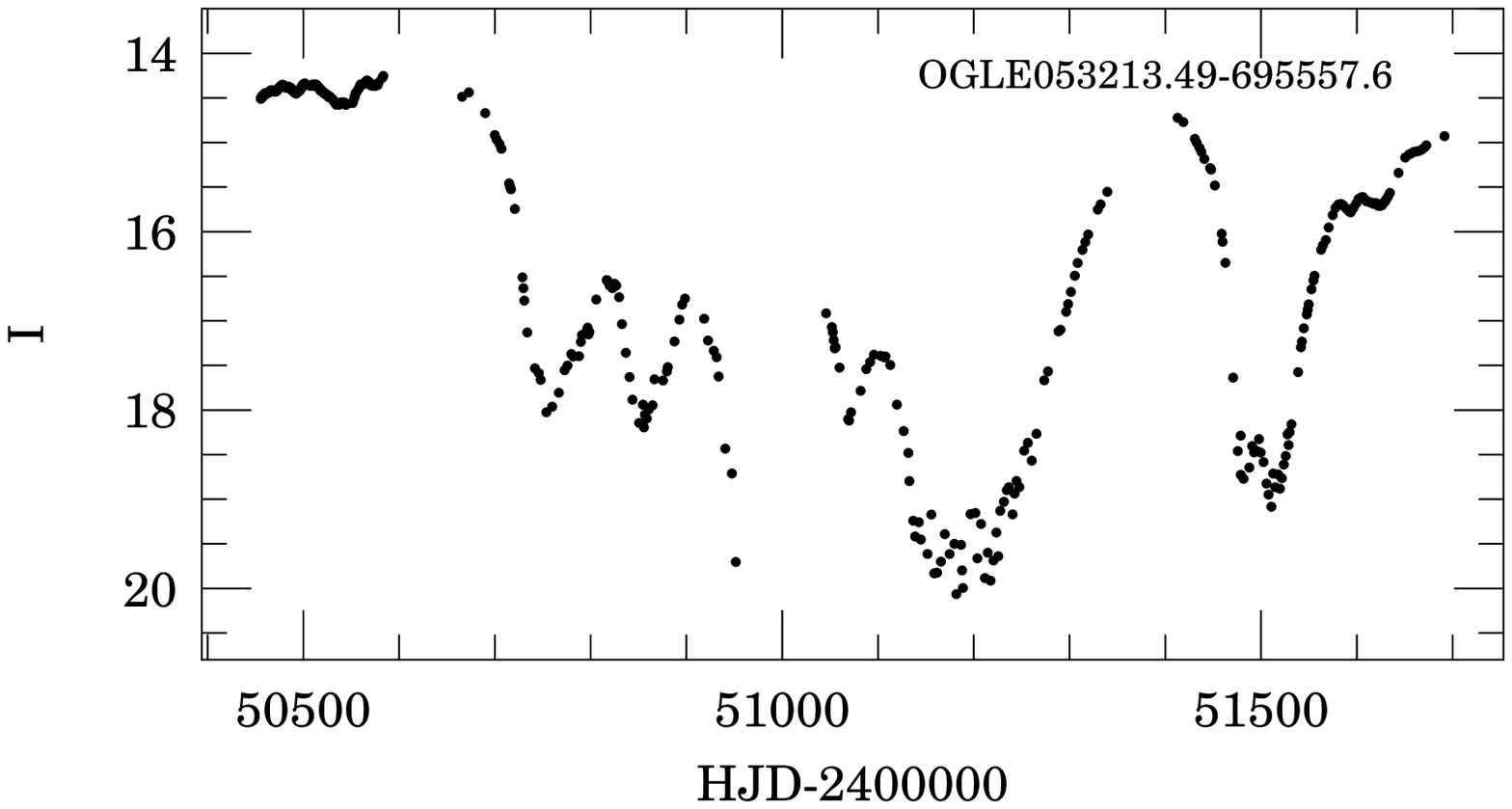}}
\FigCap{OGLE053213.49--695557.6 -- a candidate RCB variable. The light curve 
shows a quiescence period, a subsequent drop in magnitude, followed by long 
period of return to the~quiescent magnitude.} 
\end{figure}

The catalog is only weakly filtered and even though many of the artifacts were 
removed (\cf Section~7 of Paper~II) many of them are still present. To help 
the reader to find the stars that are uncertain we marked them as uncertain 
in the list of variable objects. 

Our classification criterion was very simple. For each star we calculated 
number of good measurements in flux ($N_{\rm flux}$). We did the same for 
magnitude ($N_{\rm mag}$) but only measurements inside 12--19.5~mag 
limits\footnote{The 12~mag limit is the typical saturation level for OGLE-II 
data, while the 19.5~mag limit is subjectively chosen threshold below which 
measured magnitudes become uncertain.} were taken into account. Then we 
calculated mean magnitude of the star, $\sigma$ of all measurements and number 
of measurements over ${+5\sigma}$ ($N_{5\sigma}$) and ${+10\sigma}$ limit 
($N_{10\sigma}$) above mean magnitude. The star was treated as uncertain when 
$N_{\rm mag}$ was less than $N_{\rm flux}/2$ and when $N_{5\sigma}$ and 
$N_{10\sigma}$ were less than 10 or 3 respectively. The uncertain stars are 
marked in the catalog as uncertain. We left to the reader the final judgment 
whether the variable star is real or it is only an artifact. 

The AC catalog light curves are presented in flux and magnitude units. The 
transformation to magnitude system is described in Section~4 of Paper~II. The 
linear flux light curves contain only the AC part of the signal. 

The time vector contains Heliocentric Julian date (HJD) minus 2~400~000.0 and 
the time corresponds to the beginning of the driftscan. One has to remember 
that in the driftscan mode the mid-exposure time is different for objects 
located in different parts of the image. For a given object the mid-exposure 
time can be found with accuracy to a few seconds by adding a correction 
$\Delta t$: 
$$\Delta t=(y+1024)\times 0.060776/86400~~~{\rm [days]}$$
where $y$ is the coordinate of the object in the reference image.

For convenience the DIA {\it I}-band measurements expressed in magnitudes have 
the zero point of DC signal added. They were also calibrated to standard 
system using relations derived for the OGLE data (Udalski \etal 1998a, 2000). 
For comparison purposes we also supplemented the AC part of the catalog with 
the regular OGLE {\sc DoPhot} photometry in {\it I}, {\it V} and {\it B}-band 
for objects that are best identified with the DIA variable star positions. 

The measurements of AC signal presented in the catalog are very precise. 
However this statement is not true for the DC flux measurement on a reference 
image. Depending on distance of cross-identification of a variable candidate 
from DIA with DIA {\sc {\sc DoPhot}} positions of stars we decided to use DIA 
{\sc DoPhot} DC signal or DIA DC signal. The DIA {\sc DoPhot} DC flux is 
affected by the fact that the positions of variable objects returned by DIA 
are often positions of blends of stars detected by DIA {\sc DoPhot}. The DIA 
DC flux is also affected because DIA is not modeling the star's vicinity on a 
reference frame and not removing nearby stars prior to calculating the flux. 
This makes the correct measurement of a DC signal a serious problem. Currently 
the OGLE-III phase of the OGLE project is underway. With better spatial 
resolution than available in OGLE-II, we will be able to extract more precise 
signal with both {\sc DoPhot} and DIA photometry. For details about DC flux 
measurement we refer to Section~5 in Paper~II. 

\Section{How to Use the Catalog}
The catalog is available on-line through FTP and WWW from the OGLE Internet 
archive. Here we present a brief instructions for users. 

\subsection{The Catalog Through FTP}
The catalog that can be accessed {\it via} anonymous ftp at the following addresses: 
\begin{center}
{\it ftp://bulge.princeton.edu/ogle/ogle2/dia/}\\
{\it ftp://sirius.astrouw.edu.pl/ogle/ogle2/dia/}\\
\end{center}
The catalog is placed in two subdirectories {\it dc/} and {\it ac/} containing 
data for DC and AC signal respectively. Below we summarize contents of these 
directories. 

For the DC catalog there are the following directories:
\vspace*{-6pt}
\begin{itemize}
\itemsep=-4pt
\item {\it dc/lmc/} and {\it dc/smc/} -- the DIA reference images:
16~MB gzip compressed FITS images of the whole fields,
\item {\it dc/lmc$\_$scN/} or {\it dc/smc$\_$scN/} where $N$ means the field 
number; these are the reference images stored as 256 FITS images of 
subsections; the name of a single gzip compressed file is {\it 
ref\_X\_Y.fts.gz}, where $X$ and $Y$ are the location of the subframe within a 
${4\times 64}$ partition of the full frame, 
\item {\it dc/dia$\_$dophot/} -- {\sc DoPhot} photometry on the DIA reference 
image, 
\item {\it dc/zero$\_$points/} -- magnitude zero points for all fields (by 
sections), 
\item {\it dc/tables/} -- text tables for all LMC and SMC fields (one row per 
variable, see the previous Section). 
\end{itemize}

The AC catalog contains complete photometry for all variable stars. For 
convenience we compressed archived photometry for whole fields. The file names 
are {\it lmc\_scN.tar.gz} and {\it smc\_scN.tar.gz}. Individual photometry 
files for each variable are named using our new coordinate naming convention: 
\vspace*{-6pt}
\begin{itemize}
\itemsep=-4pt
\item {\it OGLEhhmmss.ss-ddmmss.s.flux} -- AC signal in the DIA flux units,
\item {\it OGLEhhmmss.ss-ddmmss.s.mag} -- magnitude light curves, with DC 
signal and zero point added. 
\end{itemize}

The AC section of the catalog available through FTP fills approximately 0.4~GB 
of disk space and DC section approximately 1.6~GB. 

\subsection{The Catalog Through World Wide Web Page}
We also created a WWW user interface. The catalog may be updated in the 
future, but the general form of data access will remain similar. The whole 
interface is prepared in such a way that expansions and modifications do not 
influence the~availability of the catalog. The WWW catalog allows one to get a 
considerable amount of data for each suspected variable, and it is available 
at the following addresses: 
\begin{center}
{\it http://bulge.princeton.edu/\~{}ogle/ogle2/dia/}\\
{\it http://sirius.astrouw.edu.pl/\~{}ogle/ogle2/dia/}\\
\end{center}
The main WWW catalog page is divided into two frames. In the left frame there 
are links for easy browsing the remote parts of the catalog. The contents of 
the right panel depends on the current choice from the menu on the left. The 
WWW catalog has two major parts referred to as CONSTANT DATA and VARIABLE 
DATA. 

By entering the CONSTANT DATA part, one loads a map with locations of the LMC 
and SMC fields. A click inside the contour of a given field, allows to access 
corresponding data from the DC catalog. An example of a single window for 
LMC$\_$SC17 field is shown in Fig.~5. The reference frame is displayed at the 
center, with a white pane superimposed on this image. The numbers on the sides 
help to find a given subsection of the reference image. A click selects  
given subsection for a download. There are also text links to the remaining 
CONSTANT DATA as described in the FTP section. 
\begin{figure}[htb]
\vglue-5mm
\centerline{\includegraphics[width=12.5cm]{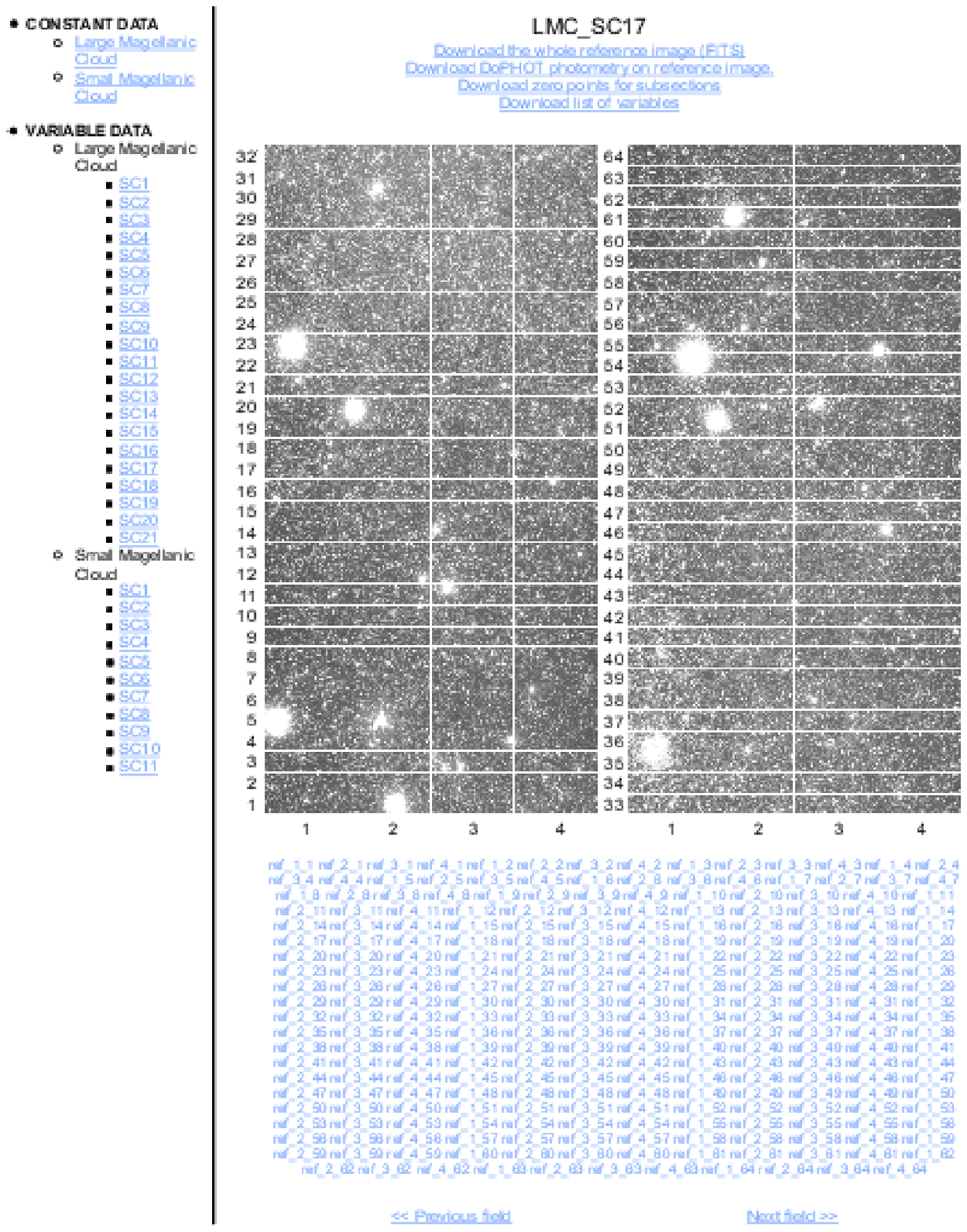}}
\FigCap{CONSTANT (DC) DATA window in the on-line catalog. The left panel 
provides links for easy navigation of the VARIABLE (AC) and CONSTANT parts of 
the catalog. The right panel displays the DIA reference image. The pane on the 
image shows ${512\times128}$ pixel subsections in two strips. The user has a 
choice of downloading the entire reference image (FITS), or any of the 
subframes separately, using either text links or clickable image sections.} 
\end{figure}

In the VARIABLE part of the WWW catalog, the user can browse lists of the all 
suspected variable objects. The table (Fig.~6) contains: variable name, $x$ 
and $y$ coordinates as returned by the DIA, {\it I}-band magnitude of the 
closest star detected by {\sc DoPhot} on the reference image and distance to 
this star. This distance is frequently larger than its expected error because 
the variable found using DIA is typically blended on the frame used for {\sc 
DoPhot} photometry.  

The name of the variable is also a link, which takes the user to a window with 
additional information about the star. Fig.~7 shows the outlook of the window 
in the VARIABLE DATA part. The table in the upper part of the window contains 
coordinates of the star and information about stars that were closely 
identified on the DIA reference frame and in the OGLE database. A finding 
chart and light curve of the object are created dynamically. If needed 
straight lines marking 12~mag and 19.5~mag limits are also plotted (see 
Section~3). The size of the finding chart (part of the deep reference image) is 
${168\times168}$ pixels, corresponding to $70\arcs\times70\arcs$ on the sky. 
The North is up and East is to the left. In the right panel, the file with 
photometry is displayed. One can choose between the DIA {\it I}-band flux and 
magnitude units, and the OGLE {\it I}, {\it V}, {\it B}-band photometry. 
\begin{figure}[p]
\centerline{\includegraphics[width=12.5cm]{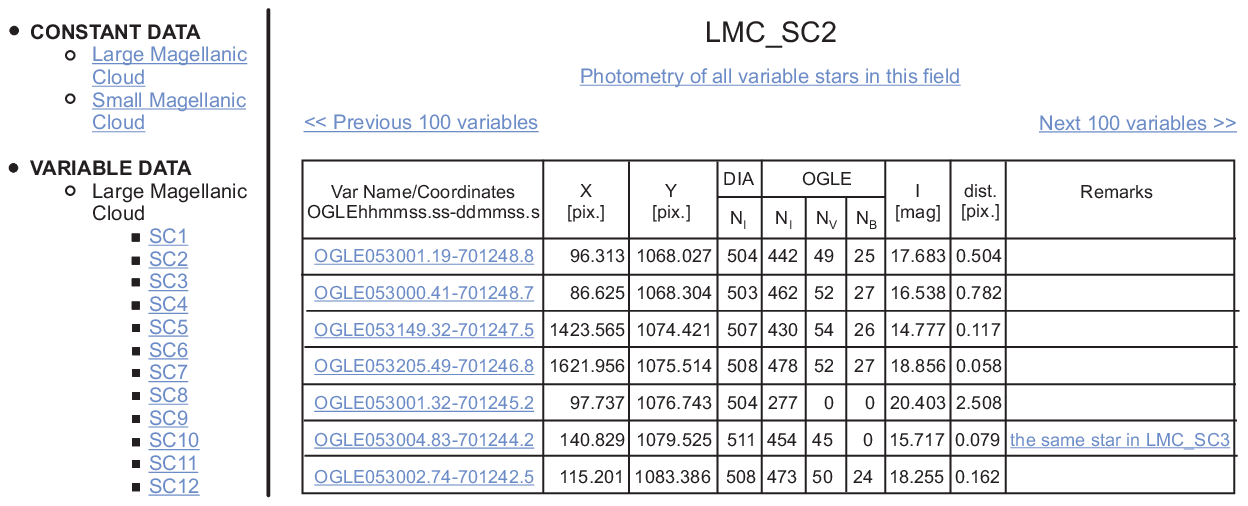}}
\FigCap{A screen shot of a star list from the AC part of the on-line catalog. 
The links to the DC and AC part of the catalog are in the left panel. In the 
right panel there is a table with a clickable list of variable stars. Each 
variable name provides a link to a new window with details about the star, see 
Fig.~7.} 
\vskip1cm
\centerline{\includegraphics[width=12.5cm]{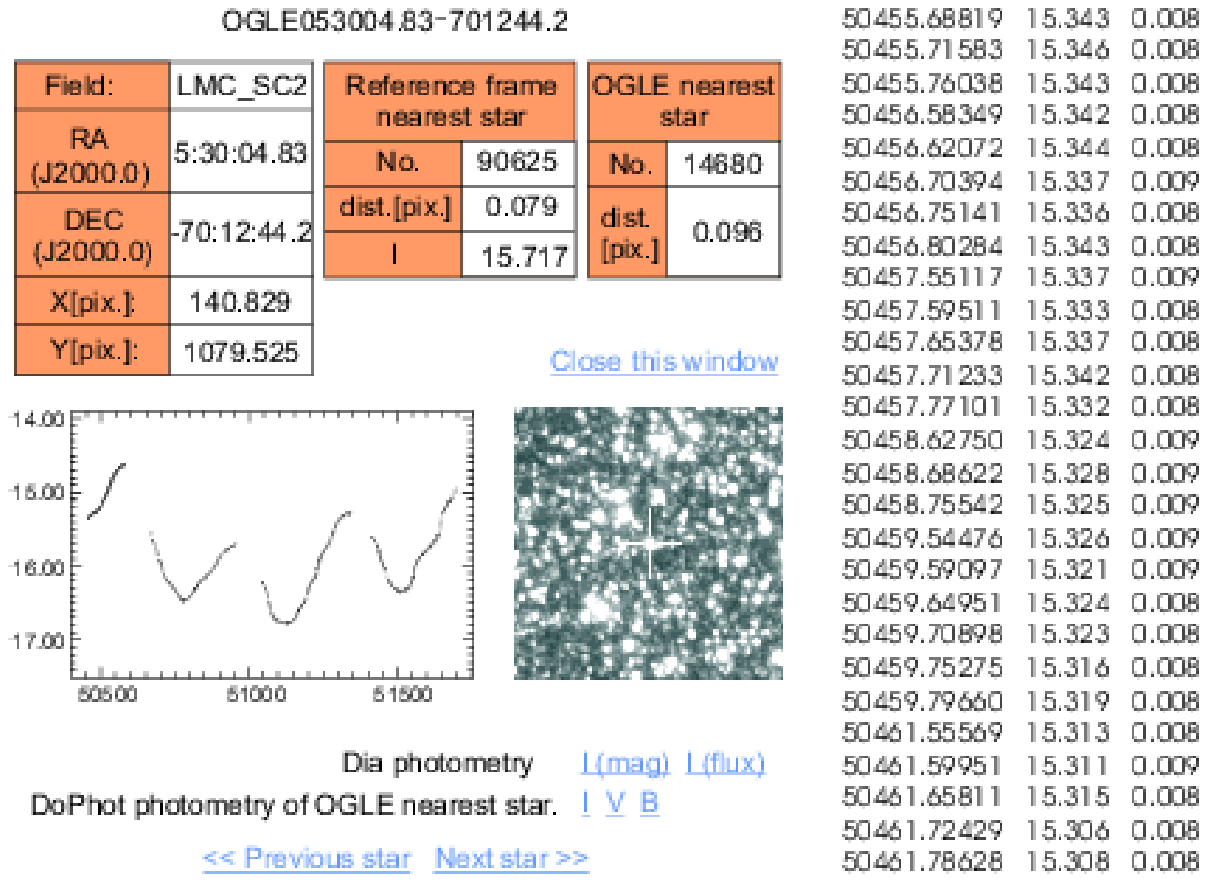}}
\FigCap{A sample window with detailed data about a single variable star. In 
the upper left corner there are tables with coordinates, mean {\it I}-band 
magnitudes etc. In the center one can see a raw light curve in magnitude 
units and a ${70\times70}$ arcsec finding chart. The right part of the window 
displays photometric data points for the corresponding star. There is 
a selection of data between the DIA photometry in magnitude, the DIA 
difference flux and the OGLE photometry of the nearest star in {\it B}, {\it 
V} and {\it I}-band.} 
\end{figure}

\Section{Summary}
The online catalog of OGLE-II candidate variables in the LMC and SMC from the 
DIA photometry contains light curves for more than 68~000 variable stars. 
Currently, it is a preliminary version which we expect to evolve towards a 
refined product, free of artifacts, more complete, with added complexity of 
scientific information like variability classes etc. The stars with high 
proper motions will be presented in a separate catalog. The main strengths of 
this work are precise differential photometry and very modest assumptions 
about included variability types, with the potential for finding new 
information on exotic objects discovered using other means or even the catalog 
itself. We encourage astronomers to make comments and propose improvements for 
the future versions. 

Users can also obtain a copy of FTP catalog (approximately 2~GB) on a DAT 
tape. The~request should be sent to Prof.\ B.\ Paczy\'nski 
(bp@astro.princeton.edu). 

\Acknow{It is a pleasure to acknowledge that this work begun when two of us 
(KZ, IS) were visiting the Department of Astrophysical Sciences at Princeton 
University, and one of us (PW) was a graduate student at that department. This 
work was partly supported by the KBN grant 5P03D~025~20 to I.\ Soszy\'nski, 
2P03D~014~18 to M.\ Kubiak and 5P03D~027~20 to K.\ \.Zebru\'n. Partial 
support was also provided by the NSF grant AST-9830314 to B. Paczy\'nski.}

\end{document}